\documentclass[conference]{IEEEtran}
\IEEEoverridecommandlockouts
\usepackage{cite}
\usepackage{amsmath,amssymb,amsfonts}
\usepackage{algorithmic}
\usepackage{graphicx}
\usepackage{textcomp}
\usepackage{xcolor}
\usepackage{bm}
\usepackage{algorithmic,algorithm}
\usepackage{multirow}
\usepackage{nidanfloat}
\newcommand{\argmin}{\mathop{\rm arg~min}\limits}
\def\BibTeX{{\rm B\kern-.05em{\sc i\kern-.025em b}\kern-.08em
    T\kern-.1667em\lower.7ex\hbox{E}\kern-.125emX}}
\begin{document}

\title{Refining Similarity Matrices to \\Cluster Attributed Networks Accurately}

\author{\IEEEauthorblockN{Yuta Yajima and Akihiro Inokuchi}
\IEEEauthorblockA{Graduate School of Science and Technology, Kwansei Gakuin University}
}

\maketitle

\begin{abstract}
As a result of the recent popularity of social networks and the increase in the number of research papers published across all fields, attributed networks---consisting of relationships between objects, such as humans and the papers, that have attributes---are becoming increasingly large. Therefore, various studies for clustering attributed networks into sub-networks are being actively conducted. 
When clustering attributed networks using spectral clustering, the clustering accuracy is strongly affected by the quality of the similarity matrices, which are input into spectral clustering and represent the similarities between pairs of objects. In this paper, we aim to increase the accuracy by refining the matrices before applying spectral clustering to them. We verify the practicability of our proposed method by comparing the accuracy of spectral clustering with similarity matrices before and after refining them.
\end{abstract}



\section{Introduction}
\label{1}

In recent years, developments in information technology have led to the accumulation of a very large amount of data \cite{Khan}, which can no longer be managed by humans.
Therefore, studies on data mining techniques to identify useful knowledge from a very large amount of data are attracting substantial attention.
Among these techniques, clustering is useful for grouping data into clusters automatically, which is difficult to perform manually.
Examples of data that have become unmanageable include attributed networks~(ANs), each of which consists of a set of objects with attributes and relationships between the objects.
Examples of ANs include social networks, which consist of relationships between humans, and citation networks, which consist of citations between documents.
When we analyze a social network, the attributes of each human are his/her characteristics, such as the number of times to tweet words, and relationships between humans include follow relationships.
Clustering an AN into groups, each containing humans with a high similarity to each other, has been actively studied \cite{Bothorel}\cite{Zhiqiang}\cite{Zhiqiang2}.

By clustering objects in an AN into groups, we can understand some properties of the AN that cannot be discerned manually; this provides various benefits.
For example, in social networks, groups are considered to be human communities; when companies use social networks to advertise their products, they can distribute advertisements to communities that may be interested in their products.
Companies expect this to increase the efficiency of advertisements.

In this paper, we use spectral clustering \cite{Luxburg} to cluster ANs. Its accuracy is strongly affected by the quality of the similarity matrices, which are necessary input for spectral clustering and represent the similarity between objects.
In this paper, we aim to increase the accuracy by refining the similarity matrices before applying spectral clustering to them.

The contributions of our paper are summarized as follows:
\begin{itemize}
  \item We propose a process to combine two different types of features of an AN, attributes and relationships, to derive a high-quality similarity matrix.
  \item We demonstrate that the refinement process that increases the quality of the similarity matrix is effective in the clustering of ANs using various real-world databases.
\end{itemize}

The remainder of this paper is organized as follows: In
Section~\ref{3}, we define ANs and introduce the clustering problem
for ANs. We also provide an overview of the refinement of
similarity matrices. In Section~\ref{4}, we propose algorithms to
cluster objects in an AN using the refinement process. In
Section~\ref{5}, we present the experimental results for real-world databases. In Section~\ref{6}, we discuss out proposed
method and related work. In Section~\ref{7}, we conclude this
paper.


\section{Clustering of Attributed Networks using the Refinement of Similarity Matrices}
\label{3}

\subsection{Clustering of Attributed Networks}
\label{3.1}

First, we define ANs and introduce the clustering problem for ANs.
An AN that consists of $n$ objects $O=\{o_1, \dots , o_n\}$ is a directed network $N=(O,A,L)$, with two features $A=(\bm{a}_1, \dots ,\bm{a}_n)$ and $L$, described as follows:

\begin{description}
\item[$A:$] Each $o_i$ has a $d$-dimensional attribute vector $\bm{a}_i=(a_{i1}, \dots ,a_{id})$ whose elements represent the object's characteristics.
\item[$L:$] Each element in $L$ is a direct relationship between objects $o_i$ and $o_j$, which represents a relationship such as a follow relationship or citation.
\end{description}

\noindent
The method proposed in this paper is applicable to any AN with attributes and direct relationships, and is not restricted to social networks and citation networks.
Given an AN and the number of clusters $c$ as input, the clustering problem for ANs is to group the set of objects $O$ into $c$ disjoint subsets $\mathcal{C}=\{C_1, \ldots, C_c\}$ called clusters, each of which consists of objects that have a high similarity to each other.

We use spectral clustering to solve the clustering problem for the ANs defined above.
The input for spectral clustering is a similarity matrix $S\in \mathbb{R}^{n\times n}$, whose $(i, j)$ element $S(i, j) \ge 0$ represents the similarity between objects $o_i$ and $o_j$.
Therefore, to perform spectral clustering, it is necessary to derive a similarity matrix $S$.
We combine the two features of an AN, $A$ and $L$, to derive a high-quality $S$.
Specifically, the higher the value in $S$, the greater the similarity between the attributes of the two objects and the stronger the direct relationship between them.
We propose the process to derive $S$ in Section~\ref{4.2}.
We also discuss the refinement process, which increases the quality of $S$, in Sections~\ref{3.2} to \ref{3.4}.

\subsection{Relationship Between Accuracy and Matrix Quality}
\label{3.2}

\begin{figure}[t]
\begin{center}
\includegraphics[width=8.1cm, height=3.4cm]{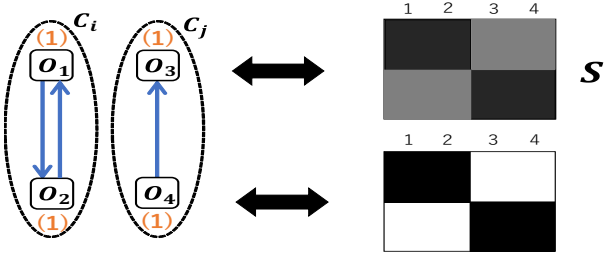}
\end{center}
\caption{An AN, low-quality similarity matrix $S$, and high-quality similarity matrix}
\vspace{-5mm}
\label{highqualityS}
\end{figure}

To explain the refinement process, we discuss the relationships between the accuracy of spectral clustering and the quality of the similarity matrices.
We consider the similarity matrix $S$ for the AN on the left-hand side of Fig.~\ref{highqualityS}, with the number of clusters $c=2$ as input.
The AN has features $A$ and $L$, where each $\bm{a}_i$ is a one-dimensional vector, written in orange, and each relationship is represented by a blue arrow. Additionally, the supervised clusters that we would like to detect from the AN are represented by dashed ellipses.
The similarity matrix $S$ shown in the lower right of Fig.~\ref{highqualityS} is said to be of high quality because the similarity between objects that belong to the same cluster $C_i$ is high and the similarity between objects that belong to different clusters, $C_i$ and $C_j$, is low.
When inputting a similarity matrix into spectral clustering as input, the higher the quality of the matrix, thef greater the probability of obtaining desirable clusters.
However, it is not easy to find the high-quality similarity matrix shown in the lower right of Fig.~\ref{highqualityS} because we cannot assume that supervised clusters are available.
Therefore, we actually derive the high-quality similarity matrix from only the information of $A$ and $L$, and the number of clusters $c$; this often results in the low-quality $S$ shown in the upper right of Fig.~\ref{highqualityS}.

In this paper, we aim to refine the low-quality similarity matrix $S$ to derive a high-quality similarity matrix, without using supervised clusters.
In the next subsection, we explain the refinement process.

\subsection{Refining the Similarity Matrices to Improve Quality}
\label{3.3}

We describe the refinement process that increases the quality of the similarity matrices without the need for supervised clusters.
To explain the process, we use a weighted undirected graph $G(S)=(V,E)$ that corresponds to a similarity matrix $S$.
In the graph, each vertex $v_i \in V$ corresponds to an object $o_i$, and each weight of an edge $e_{ij} \in E$ between $v_i$ and $v_j$ corresponds to $S(i,j)>0$.
We use $G(S)$ to explain the refinement of the similarity matrix $S$ because $S$ and $G(S)$ are equivalent in terms of representing ANs similarity relations.

The {\it refinement} of a similarity matrix $S$ aims to optimize $S$ to $S^c$ so that $G(S^c)$ consists of $c$ connected components to increase the quality of $S$.
To verify that the quality of the similarity matrix whose corresponding graph consists of $c$ connected components is high, we compare the refinement process with the $k$-nearest neighbor graph~($k$-nng) process \cite{Luxburg}, which is one of the common process for reconstructing graphs.

\begin{figure}[t] 
\begin{center}
\includegraphics[width=8.8cm, height=5cm]{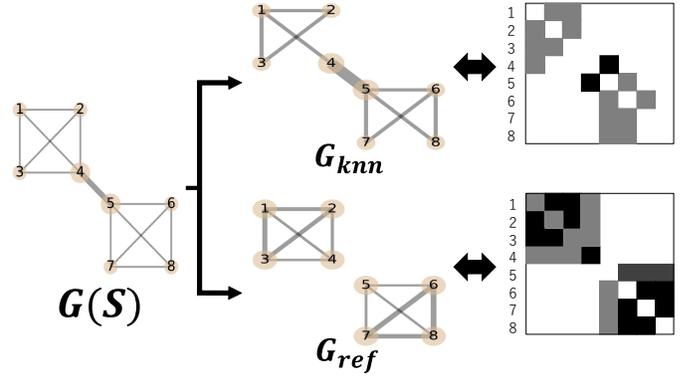}
\end{center}
\vspace{-3.5mm}
\caption{$k$-nng~$(k=2)$ process, which reconstructs $G(S)$ as $G_{knn}$, and the refinement process, which reconstructs $G(S)$ as $G_{ref}$}
\label{knnvsrefine}
\end{figure}

The left-hand side of Fig.~\ref{knnvsrefine} shows a graph $G(S)$ for an unrefined similarity matrix $S$. If we assume that the desirable number of clusters $c$ is 2, the graph $G(S)$ should be divided into two connected components that consist of $\{v_1,v_2,v_3,v_4\}$ and $\{v_5,v_6,v_7,v_8\}$.
In this scenario, the graph $G(S)$ is reconstructed as $G_{knn}$ using the $k$-nng process, and as $G_{ref}$ using the refinement process, as shown in Fig.~\ref{knnvsrefine}.
The main difference between $G_{knn}$ and $G_{ref}$ is that the edge $e_{45}$ is present in $G_{knn}$ but not in $G_{ref}$.
As a consequence of cutting the edge $e_{45}$ using the refinement process, the graph $G(S)$ is reconstructed as $G_{ref}$, which has the desired two connected components.

We now explain the reason for the difference between $G_{knn}$ and $G_{ref}$.
The $k$-nng method \cite{Luxburg} leaves only edges $\{e_{ij}\}$ if $v_j$ is one of the $k$ neighbors of $v_i$ with the largest weight or $v_i$ is one of the $k$ neighbors of $v_j$ with the largest weight.
In Fig.~\ref{knnvsrefine}, the edge $e_{45}$, which has the largest weight among the edges of both $v_4$ and $v_5$, is left in $G_{knn}$.
By contrast, the refinement process reconstructs the graph $G(S)$ as $G_{ref}$, which consists of two~$(=c)$ connected components.
Therefore, it is possible to cut $e_{45}$ to reconstruct two components, $\{v_1,v_2,v_3,v_4\}$ and $\{v_5,v_6,v_7,v_8\}$, each of whose vertices are tightly connected to each other.
The refinement process increases the similarity between vertices in each cluster that we would like to detect, and reduces the similarity between vertices that belong to different clusters, compared with the $k$-nng method.
Thus, the refinement process increases the quality of the similarity matrix without using supervised clusters.
In fact, the similarity matrix for $G_{ref}$ derived using our refinement method is almost the same as the high-quality similarity matrix shown in the lower right of Fig.~\ref{highqualityS}.

\begin{figure}[t]
\begin{center}
\includegraphics[width=8.1cm, height=3.8cm]{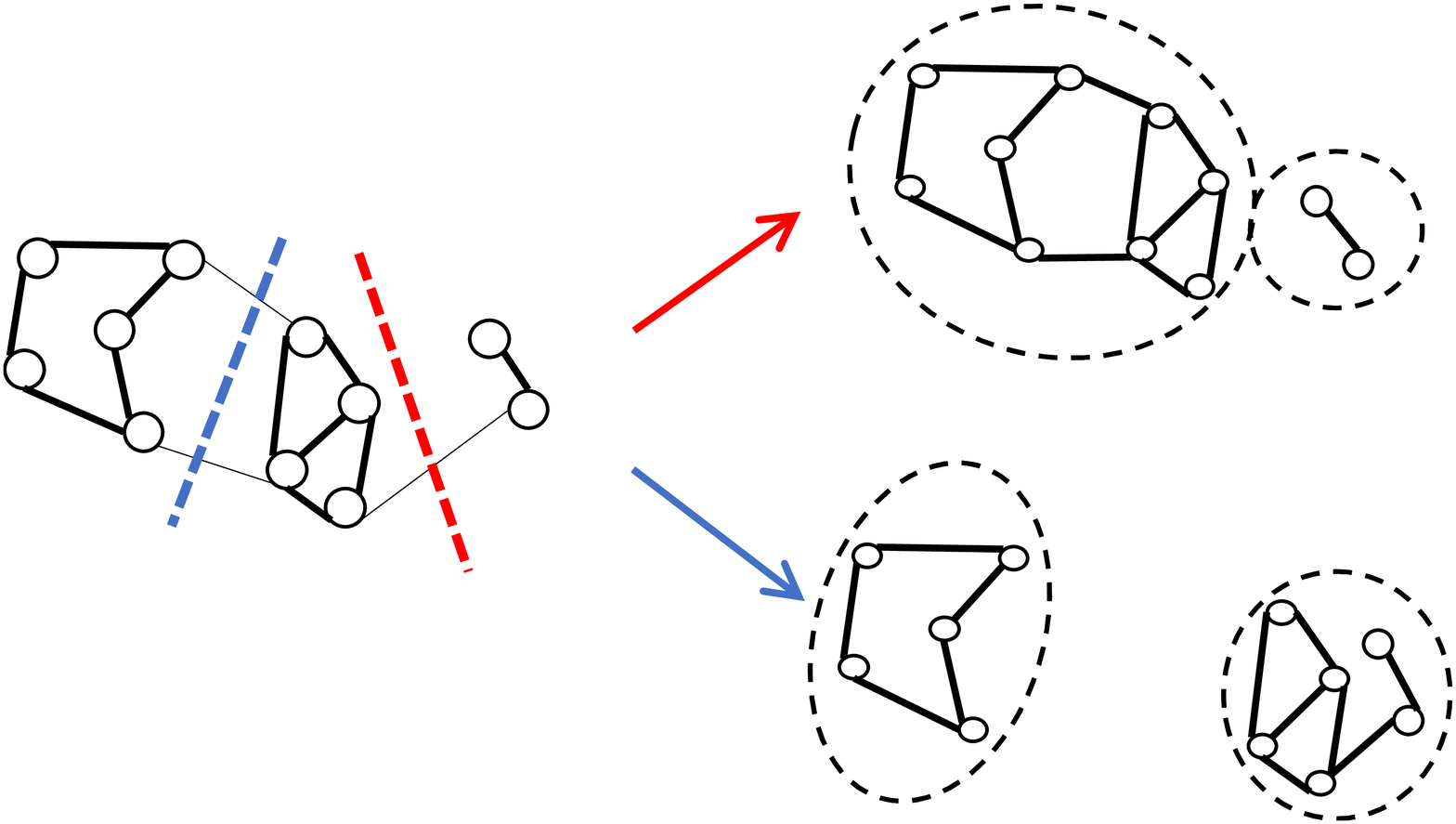}
\end{center}
\caption{Refinement with a low mean silhouette score indicated by the red
arrow and refinement with a high mean silhouette score indicated by the blue
arrow.}
\vspace{-5mm}
\label{req1}
\end{figure}

In addition to the above refinement process, our study aims to cluster ANs to achieve a high mean silhouette score.
The silhouette score is one of the most popular and effective internal measures for the evaluation of clustering validity without using supervised clusters \cite{FWang}.
The mean silhouette score is high when the sizes of the obtained clusters are almost the same and the clusters are separate from each other.
Obtaining clusters that have a high mean silhouette score prevents the refinement process from constructing connected components that are too small, that is, that consist of only a few vertices, and from obtaining them as clusters, as shown in Fig.~\ref{req1}.
In Section~\ref{4.3}, we propose a process to obtain clusters that have a high mean silhouette score.

\subsection{Application of the Refinement Process to ANs}
\label{3.4}

\begin{figure}[b]
  \begin{center}
    \begin{tabular}{c}
      \begin{minipage}{0.48\hsize}
        \begin{center}
          \includegraphics[clip, width=4cm, height=4cm]{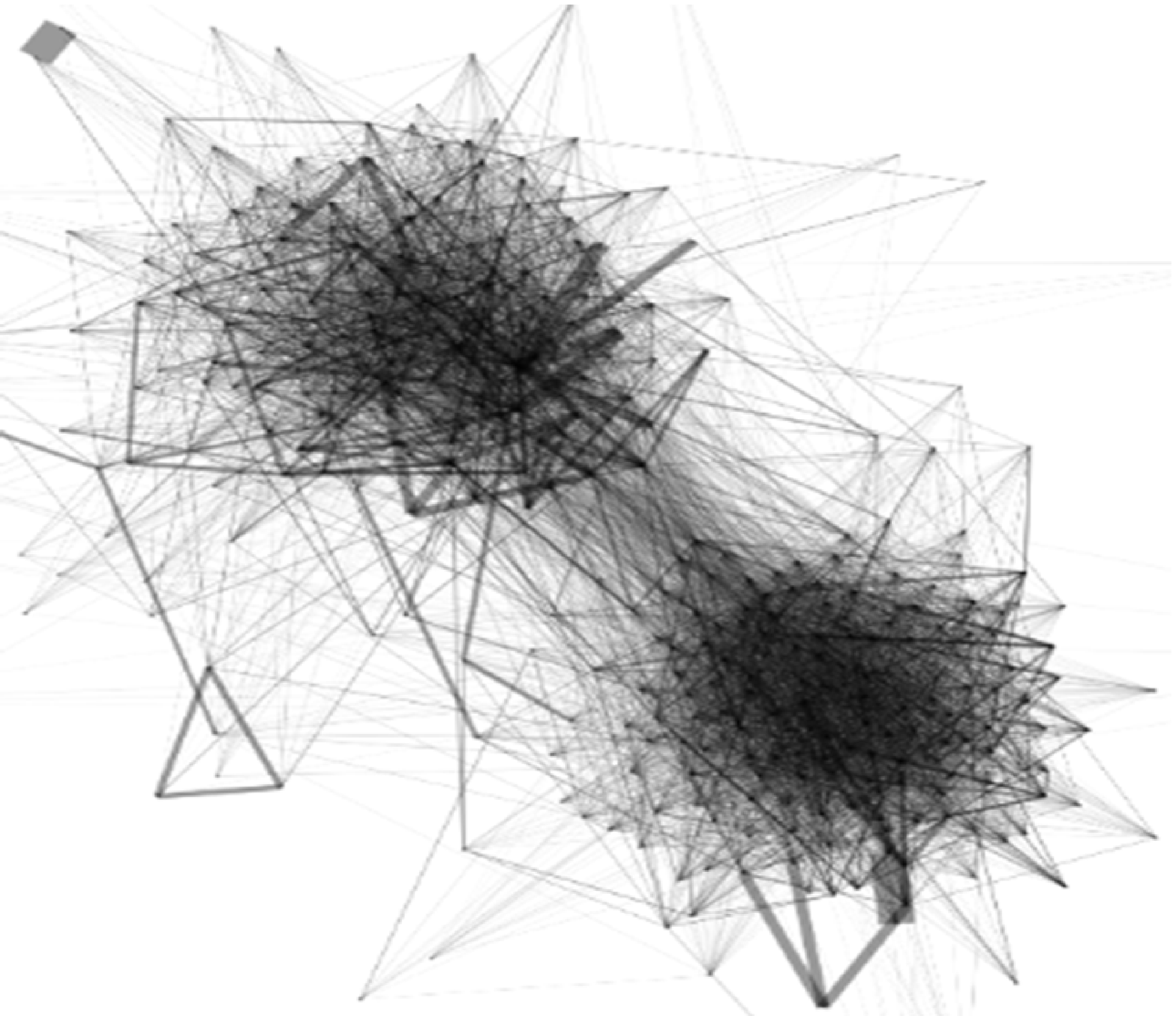}
          \hspace{+10mm} politicsuk
        \end{center}
      \end{minipage}

      \begin{minipage}{0.48\hsize}
        \begin{center}
          \includegraphics[clip, width=4cm, height=4cm]{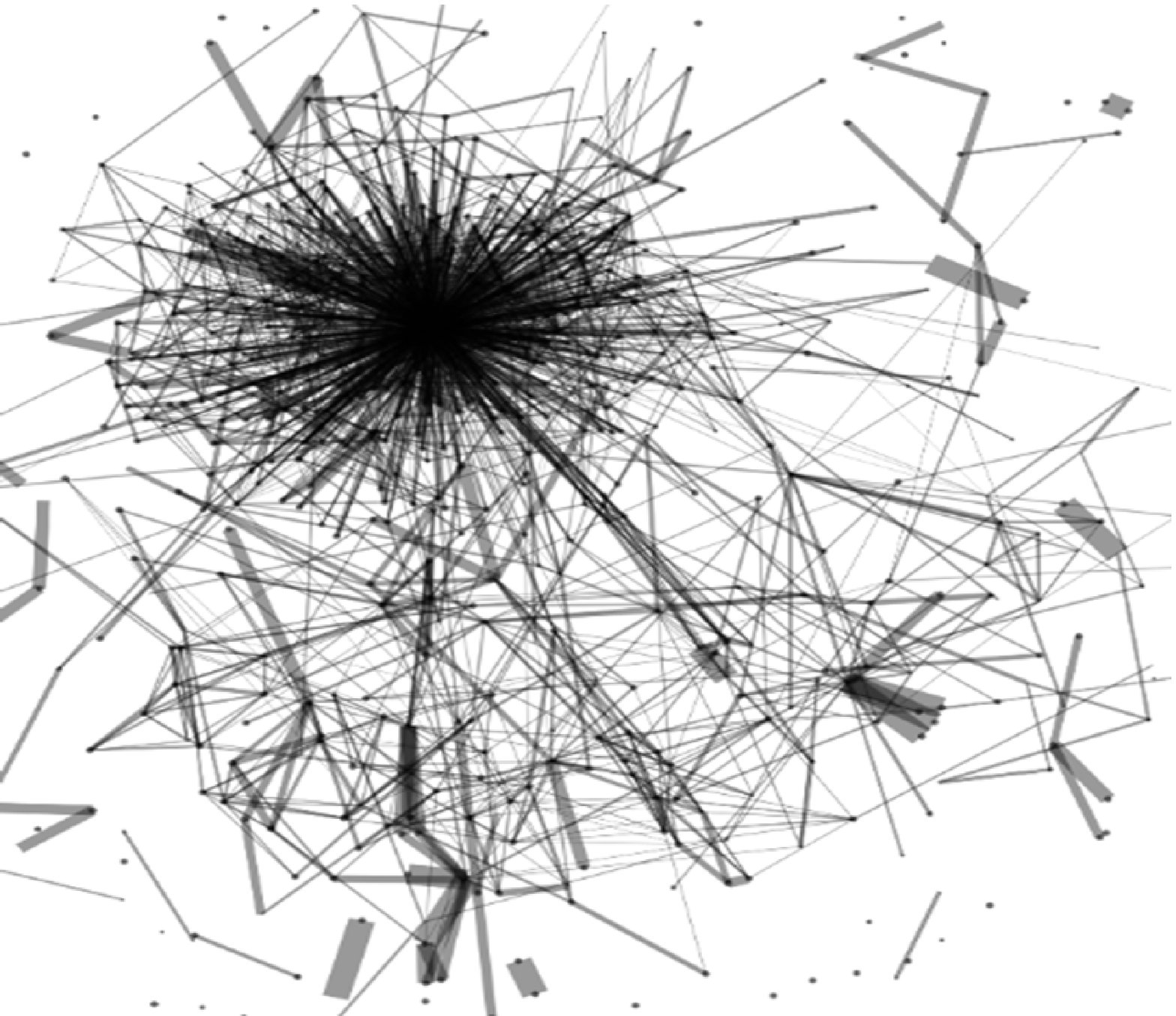}
          \hspace{+20mm} cora
        \end{center}
      \end{minipage}

    \end{tabular}
    \vspace{+1.5mm}
    \caption{Graphs generated from parts of a social network and a citation network}
    \label{visualization}
  \end{center}
\end{figure}

We expect that the refinement will increase the quality of the
similarity matrices for many ANs in real-world databases. For
example, Fig. 4 shows graphs for similarity matrices generated
from parts of a social network called politics-uk\footnote[1]{http://mlg.ucd.ie/aggregation/index.html} and a citation network called cora\footnote[2]{https://linqs.soe.ucsc.edu/data}. In the figure, the similarity between
objects is represented by the thickness of the edge. In Fig.~\ref{visualization},
the vertices in each cluster, in the lower right and upper left
of each of the graphs, are densely connected by thick edges.
However, the vertices in different clusters are also connected
by thick edges. For the graphs, the refinement process is
expected to reduce the weights of the edges between clusters
and increase the weights of the edges within each cluster to
increase the quality of the similarity matrices.

The authors of an existing clustering method called SClump \cite{Li} proposed the above refinement process.
However, SClump is applic able to networks in which ``each object has only a single attribute of a discrete value but not a continuous value,'' which is not true of our ANs. 
Therefore, SClump has the drawback that the refinement process, which was proposed in \cite{Li}, cannot be applicable to the clustering of ANs. In this paper, we propose a method for clustering ANs using a refinement process by extending SClump.


\section{Proposed Method}
\label{4}

In Sections~\ref{3.2} to \ref{3.4}, we presented an overview of the refinement of similarity matrices and described SClump's drawback: its refinement process is not applicable to the ANs considered in this paper.
In this section, we propose a method for clustering ANs using a refinement process by extending SClump.

When refining a similarity matrix $S$ to $S^c$, if the quality of the unrefined $S$ is high, the quality of the refined $S^c$ will also be high.
Hence, it is very important not only to refine $S$ to $S^c$ but also to derive $S$ from an AN.
Therefore, in Section~\ref{4.2}, we also propose a process for deriving $S$.
Then in Section~\ref{4.3}, we provide the details for an approach to refine $S$ to a high-quality $S^c$.

\subsection{Process to Derive the Similarity Matrix}
\label{4.2}

We now explain the process to derive the similarity matrix $S$ from two features---the attributes $A$ and relationships $L$---of the AN given as input.
Let the similarity matrix derived from $A$ be $S_A \in \mathbb{R}^{n \times n}$ and the similarity matrix derived from $L$ be $S_L \in \mathbb{R}^{n \times n}$.
We derive $S$ as the weighted sum of $S_A$ and $S_L$ with a weight vector $\bm{\lambda}$.
We explain the process to derive $S_A$ and $S_L$ as follows\footnote[3]{In the proposed process, each similarity matrix, $S_A$ and $S_L$, is the weighted sum of several matrices, and we refine the similarity matrix $S$, which increases the quality of $S$, by learning the weights from the input.}:

\vspace{+1mm}
\subsubsection{Similarity Matrix $S_A$ Derived from $A$}

To derive the similarity matrix $S_A$ from $A$, we calculate the similarity between $o_i$ and $o_j$ from the $k$-th elements of $\bm{a}_i$ and $\bm{a}_j$ as follows:

\vspace{-2mm}
\begin{eqnarray}
S_{Ak}(i,j)=\left\{ \begin{array}{ll}
0 ~~~~~~~~~~~~~~~~~~~~~~~~~ \mbox{if} \ i = j, \\
\exp \left[ -\frac{( a_{ik}-a_{jk} )^2} {2\sigma^2} \right] ~~~ \rm{otherwise.}\\
\end{array} \right.
\label{s1}
\end{eqnarray}

\noindent
We derive $d$ similarity matrices $\{ S_{Ak} \in \mathbb{R}^{n \times n} ~| ~1 \le k \le d \}$ from the $d$-dimensional vectors $\bm{a}_i$, and let the weighted sum of the $d$ similarity matrices be $S_A$.
The calculation in the refinement process may be unable to terminate in a reasonable time when $d$ is too large.
Therefore, we reduce the number of dimensions of $\bm{a}_i$ using principal component analysis~(PCA), and update each $\bm{a}_i$ to a low-dimensional vector in the preprocessing phase\footnote[4]{The contribution rates of PCA are used to initialize the weights to derive $S_A$ from $\{S_1, \dots, S_d\}$; this is described on p.~5.}.

\vspace{+1mm}
\subsubsection{Similarity Matrix $S_L$ Derived from $L$}

We calculate each similarity in $S_L$ from the direct relationships in $L$.
When calculating similarity, we take account of the direct relationship between two objects and the transitive relationships between them; this is accomplished by considering the shortest paths~(SPs) of the form $o_i \rightarrow \cdots \rightarrow o_j$.

\begin{figure}[t]
\begin{center}
\includegraphics[width=8.4cm, height=3.8cm]{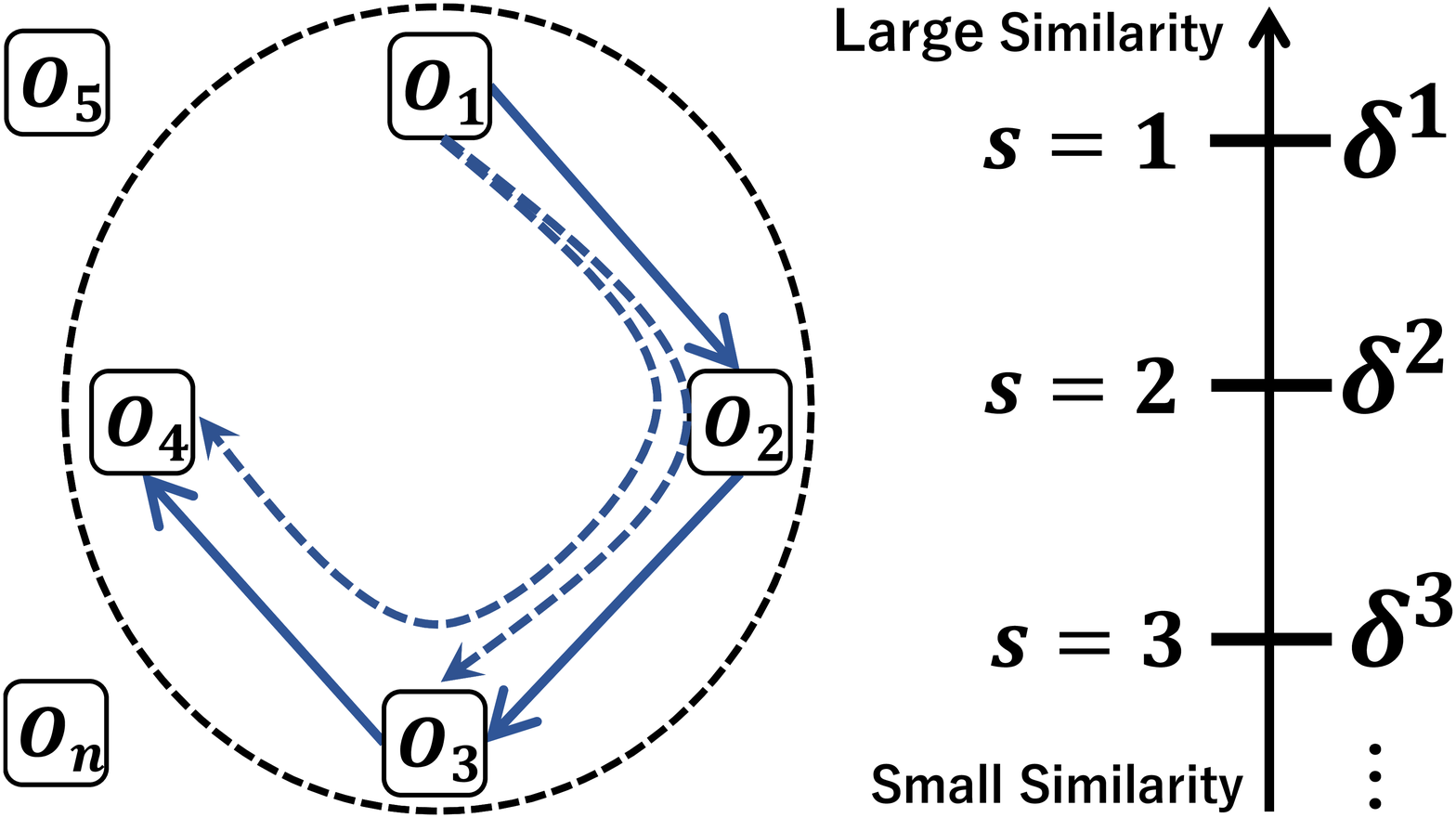}
\end{center}
\vspace{-3mm}
\caption{AN and the similarity between $(o_i, o_j)$, depending on the length $s$ of an SP}
\label{s2_trans}
\end{figure}

We illustrate the necessity of transitive relationships using an AN in Fig.~\ref{s2_trans}. The solid blue arrows in Fig.~\ref{s2_trans} represent the direct relationships between objects and the dotted blue arrows represent the transitive relationships, which correspond to the SPs between the objects.
Specifically, we regard the SPs $o_i \rightarrow \cdots \rightarrow o_j$, whose length $s$ is less than or equal to $\theta$, as the transitive relationships between $o_i$ and $o_j$. (Hereinafter, we write ``between $o_i$ and $o_j$'' as ``between $(o_i, o_j)$.'')
When $\theta$ is 3, as shown in Fig.~\ref{s2_trans}, there is an SP $o_1 \rightarrow o_2 \rightarrow o_3 \rightarrow o_4$ between $(o_1, o_4)$.
Therefore, as shown in the dashed ellipse in Fig.~\ref{s2_trans}, transitive relationships densify the cluster $\{o_1, o_2, o_3, o_4\}$ that we would like to detect. We expect that this densification will increase the quality of $S_L$ derived from the dense cluster.
Furthermore, when calculating similarities, we would like to satisfy the following requirement.
\begin{description}
\item[Req. 1 :] ~~~The shorter the SP between $(o_i, o_j)$, the larger \\~~~ the value to which we should set the similarity \\~~~between $(o_i, o_j)$.
\end{description}

Requirement~1 is natural because direct relationships in an AN represent stronger connections than transitive relationships.
To satisfy Req.~1, when there is an SP of length $s$ between $(o_i, o_j)$, we set the similarity between the objects to be $\delta^s~(0<\delta<1)$, as shown on the right-hand side of Fig.~\ref{s2_trans}.

\begin{figure*}[t]
  \begin{center}
    \begin{tabular}{c}
      \begin{minipage}{0.48\hsize}
        \begin{center}
          \includegraphics[clip, width=8cm, height=3.5cm]{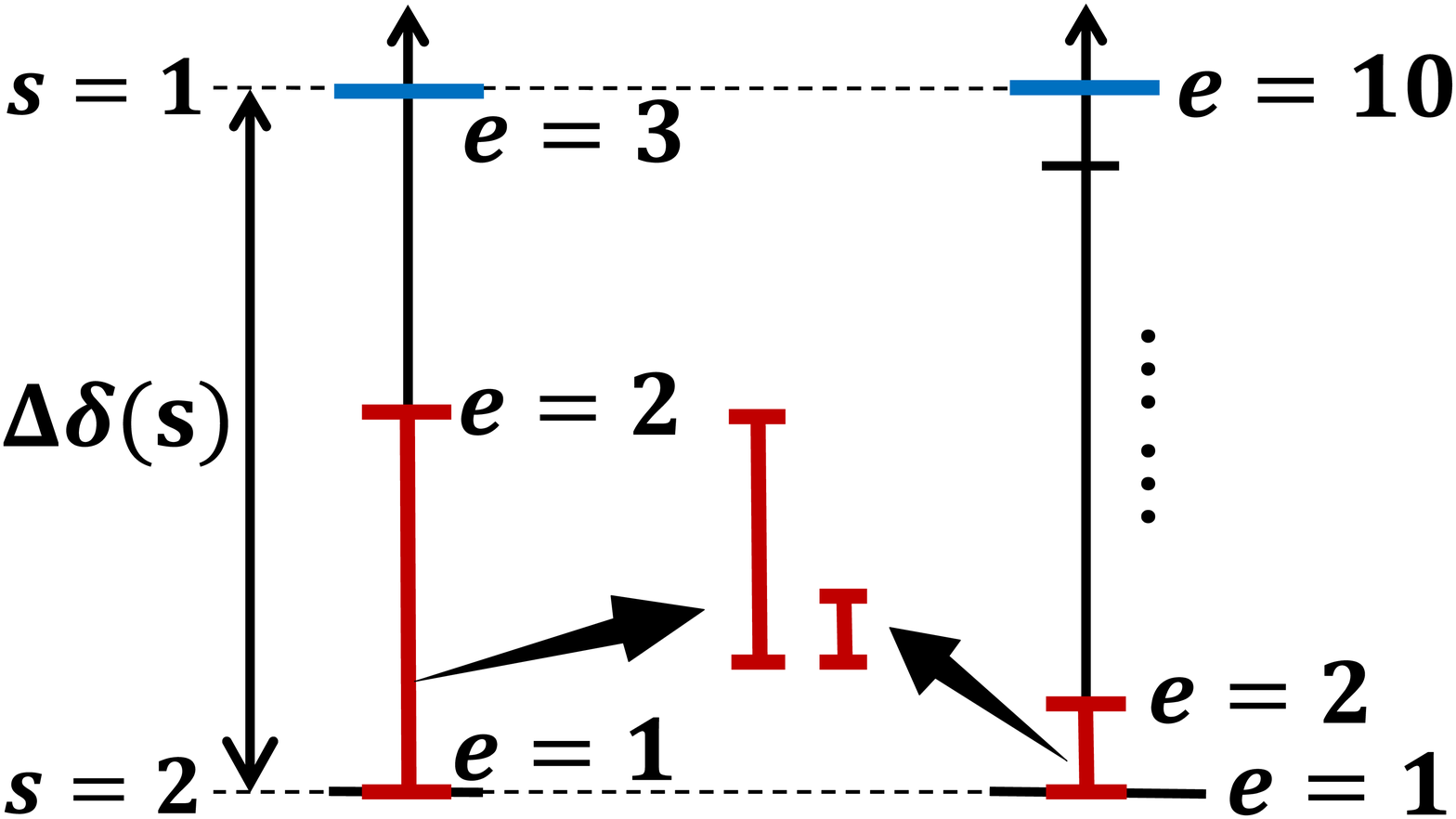}
          \hspace{+10mm} (a) Similarities to satisfy Req.~2a
        \end{center}
      \end{minipage}

      \begin{minipage}{0.48\hsize}
        \begin{center}
          \includegraphics[clip, width=7cm, height=3.5cm]{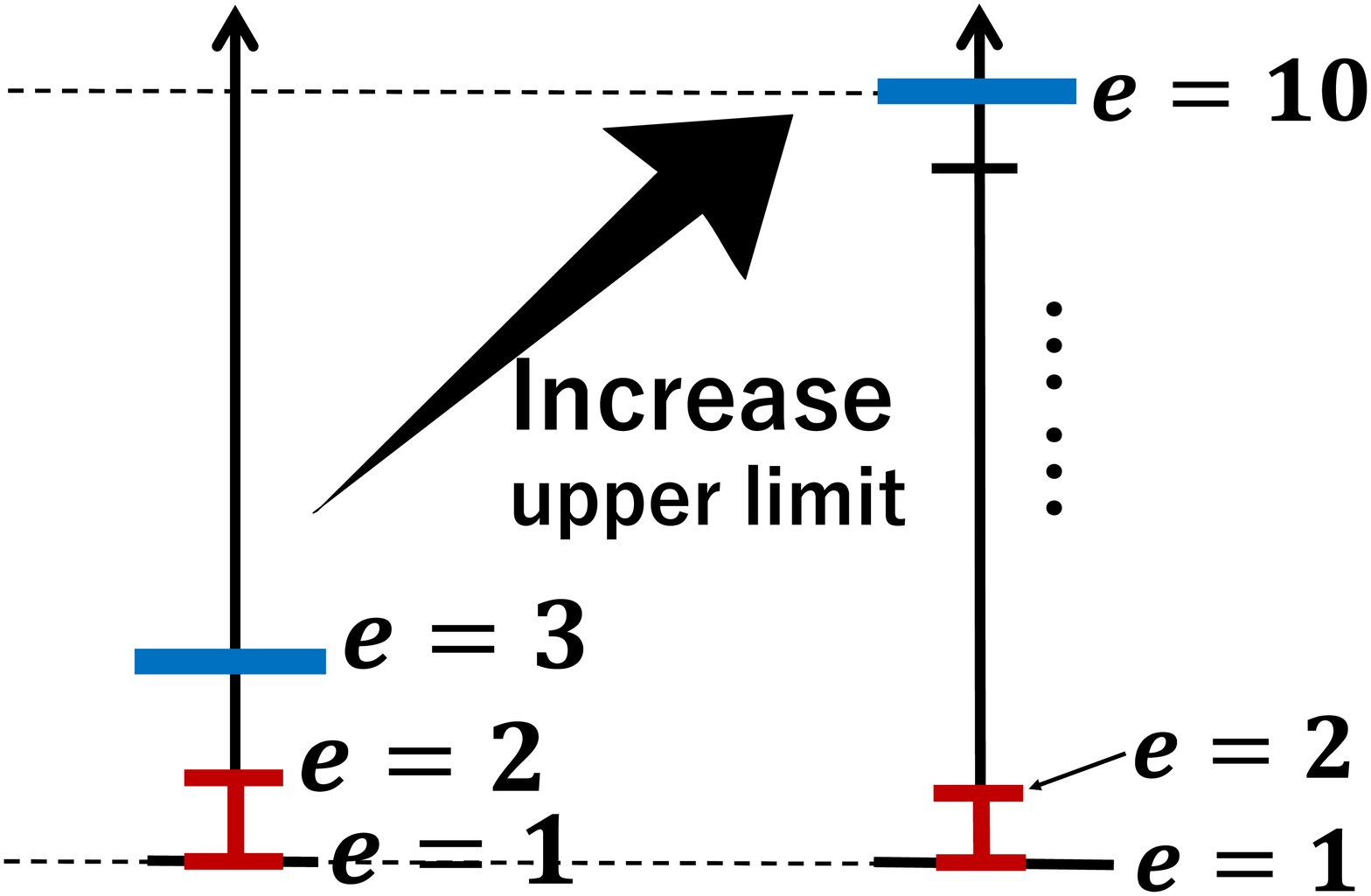}
          \hspace{+20mm} (b) Similarities to satisfy Req.~2b
        \end{center}
      \end{minipage}

    \end{tabular}
    \vspace{+1.5mm}
    \caption{Similarity between $(o_1, o_2)$ and between $(o_3, o_4)$ in Fig.~\ref{ANs_withL}}
    \label{rate}
  \end{center}
\end{figure*}

\begin{figure}[t]
\begin{center}
\includegraphics[width=8cm, height=3.6cm]{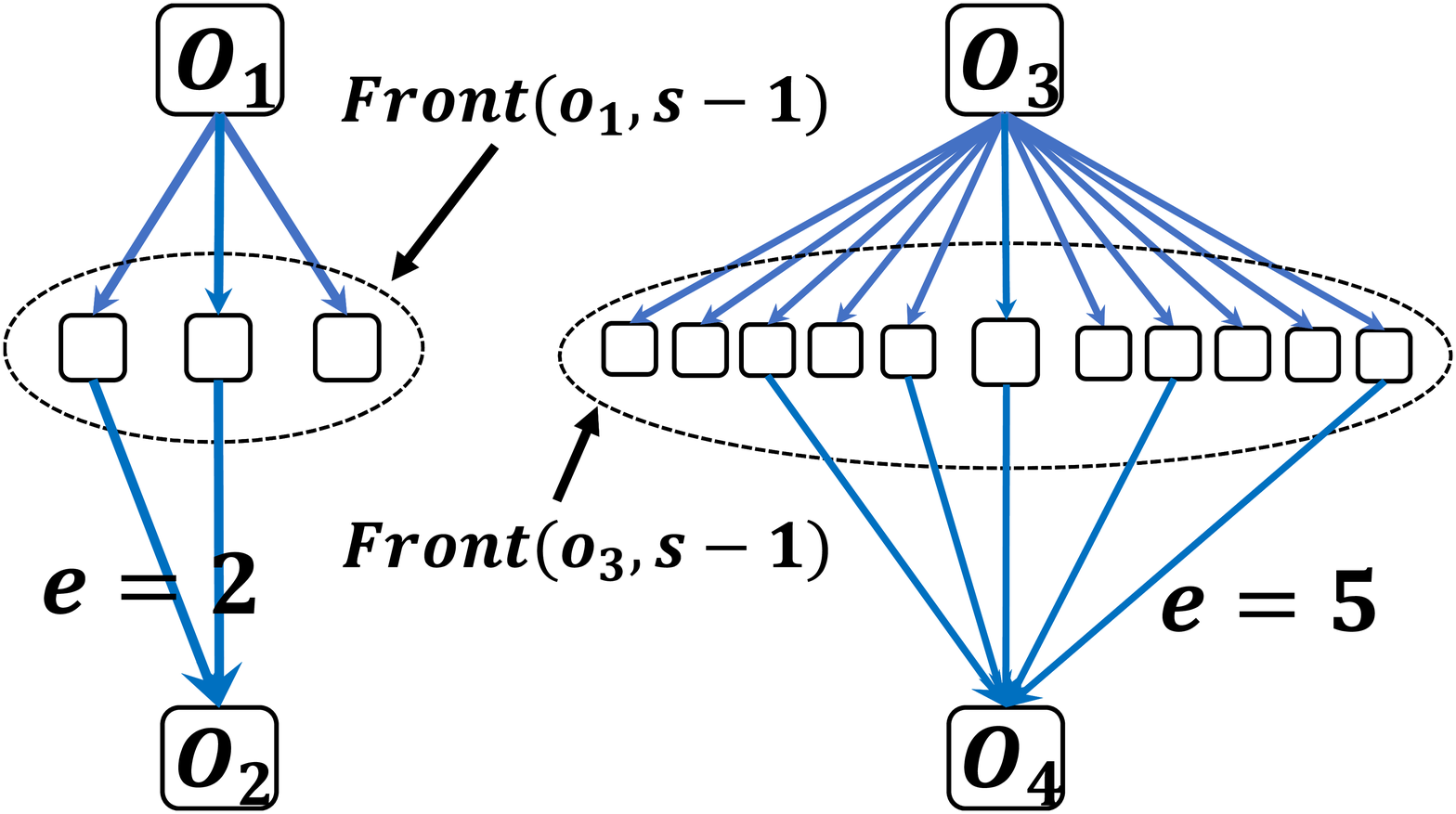}
\end{center}
\vspace{-2mm}
\caption{Meaning of $e$ and ${\it Front}(o_i, s-1)$ in an example AN}
\label{ANs_withL}
\end{figure}

We also consider other requirements for the calculation of similarities in $S_L$. In the following explanation, we use the objects $\{o_1, o_2, o_3, o_4\}$ in Fig.~\ref{ANs_withL}.
In contrast to the example in Fig.~\ref{s2_trans}, the number of SPs of length $s$ between $(o_1, o_2)$ and the number of SPs of length $s$ between $(o_3, o_4)$ are larger than 1.
In such cases, the similarities should be calculated depending not only on the length $s$ of an SP but also the number $e$ of SPs.
Moreover, when regarding objects as users, because $o_1$ follows fewer users than $o_3$, $o_1$ is considered to follow closer friends than $o_3$.
It is unlikely that the SPs between $(o_1, o_2)$ via $o_1$'s close friends represent the same strength of connections as the SPs between $(o_3, o_4)$.
Therefore, we also wish to satisfy the following requirement.

\begin{description}
\item[Req. 2 :] ~~~We wish to calculate similarities that increase \\~~~ depending on both the length $s$ and number \\~~~ $e$ of SPs.
\end{description}

\vspace{+2mm}
\subsubsection*{Similarity to Satisfy Req.~2a}

We define ${\it Front}(o_i, s-1)=R(o_i, s-1) \backslash R(o_i, s-2)$, where $R(o_i, s)$ is the set of objects reachable from $o_i$ within $s$ steps. Each SP between $(o_1, o_2)$ via $o_1$'s close friend in ${\it Front}(o_1, s-1)$ is considered to be a stronger connection than each SP between $(o_3, o_4)$ via $o_3$'s friend in ${\it Front}(o_3, s-1)$.
Therefore, as shown by the red line segments in Fig.~\ref{rate}(a), the similarity between $(o_1, o_2)$ should be greater than the similarity between $(o_3, o_4)$, even if both the $e$ between $(o_1, o_2)$ and the $e$ between $(o_3, o_4)$ increase by 1.
The strength of the connection of each SP between $(o_i, o_j)$ depends on the cardinality of ${\it Front}(o_i, s-1)$. Therefore, to satisfy Req.~2, we propose a new function to evaluate the strength of the connection:

\vspace{-2mm}
\begin{eqnarray}
S_{La}(i,j)=\left\{ \begin{array}{ll}
\delta^{s} + (\delta^{s-1}-\delta^{s}) \times \frac{e-1} {|{\it Front}(o_i, s-1)|-1} \\~~~~~~~
	\mbox{if} \ i \ne j \wedge e\ge1 \wedge s \le \theta, \\
0 ~~~~~\rm{otherwise.}
\end{array} \right.
\label{s2}
\end{eqnarray}

\noindent
The first and second terms in Eq.~(\ref{s2}) satisfy Reqs.~1~and~2, respectively. If $i \ne j \wedge e\ge 1 \wedge s \le \theta$, $S_{La}$ ranges from $\delta^{s-1}$ to $\delta^{s}$ because $\frac{e-1}{|{\it Front}(o_i, s-1)|-1}$ is the proportion of $\delta^{s-1}-\delta^{s}$ and ranges from 0 to 1.
When calculating the similarity between $(o_1, o_2)$, because $|{\it Front}(o_1, s-1)|-1 =2$, the denominator is 2.
By contrast, when calculating the similarity between $(o_3, o_4)$, the denominator is 9, which decreases $S_{La}(i,j)$.
Therefore, as shown by the red line segments in Fig.~\ref{rate}(a), Eq.~(\ref{s2}) defines a larger similarity between $(o_1, o_2)$ than between $(o_3, o_4)$, even if both $e$ between $(o_1, o_2)$ and $e$ between $(o_3, o_4)$ increase by~1.

\vspace{+1mm}
\subsubsection*{Similarity to Satisfy Req.~2b}
When all the objects in ${\it Front}(o_1, s-1)$ and ${\it Front}(o_3, s-1)$ are connected to $o_2$ and $o_4$, respectively, the latter is considered to be a stronger connection because the number of SPs between $(o_3, o_4)$ is larger than the number between $(o_1, o_2)$.
Therefore, as shown by the blue line segments in Fig.~\ref{rate}(b), we want the similarity between $(o_3, o_4)$ to be greater than the similarity between $(o_1, o_2)$ when the number of SPs between $(o_1, o_2)$ and the number between $(o_3, o_4)$ reach the maximum values under the fixed ${\it Front}(o_i, s-1)$.
We therefore propose another function for the case where $i \ne j \wedge e\ge 1 \wedge s\le \theta$ of Eq.~(\ref{s2}).

\begin{eqnarray}
S_{Lb}(i,j)=\delta^{s} + (\delta^{s-1}-\delta^{s}) \times \frac{e-1}{{\it Front(s-1)}_{max}},
\label{s2_}
\end{eqnarray}

\noindent
where ${\it Front(s-1)}_{max}=\max_{o_i \in O}|{\it Front}(o_i, s-1)|$ is constant for any two objects.
Therefore, as shown by the blue line segments in Fig.~\ref{rate}(b), when $e$ reaches the maximum value under the fixed ${\it Front(s-1)}_{max}$, the similarity between $(o_3, o_4)$ is larger than the similarity between $(o_1, o_2)$.

Because ANs are directed networks, the similarity matrices $S_{La}$ and $S_{Lb}$ are not symmetric. Therefore, we update them as $S_{La}\leftarrow (S_{La}+S_{La}^T)/2$ and $S_{Lb} \leftarrow (S_{Lb}+S_{Lb}^T)/2$ to symmetrize them.

Having obtained $\{S_{A1}, \ldots, S_{Ad} \}$, $S_{La}$, and $S_{Lb}$, we compute the weighted sum of these matrices with a weight vector $\bm{\lambda}$ to derive a similarity matrix $S$.
We initialize the $i$-th element of $\bm{\lambda}$ with the contribution ratio $p_i$ for the $i$-th principal component, as follows:

\begin{align}
\lambda_i &= \begin{cases}
			0.5 \times \frac{p_i}{\sum_{j=1}^{d}p_j} & \mbox{if}				~~ 1 \le i \le d, \\
			0.25 & \rm{otherwise.}
	         \end{cases} \nonumber \\
S &= \sum_{k=1}^{d}\lambda_k S_{Ak} + \lambda_{d+1}S_{La} +\lambda_{d+2}S_{Lb}.
\label{eq_ini_HAN}
\end{align}

\vspace{+2mm}
\subsection{Refinement Process with a Weight Adjustment}
\label{4.3}

\begin{figure}[b]
\begin{center}
\includegraphics[width=8.8cm, height=4.8cm]{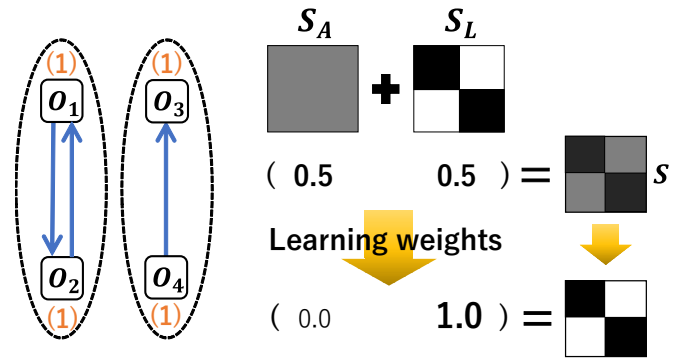}
\end{center}
\vspace{-3mm}
\caption{AN, and similarity matrices derived according to weights}
\label{lambda}
\end{figure}

We now discuss the refinement of $S$, which is the topic of this study.
We explained that we compute the weighted sum of $S_A$ and $S_L$ to derive $S$.
The left-hand side of Fig.~\ref{lambda} shows an AN and the dashed ellipses represent supervised clusters.
Because all the attribute values of the objects are 1, $S_A$, based on the attribute values, is low quality.
By contrast, because the relationships between the objects exist only in the clusters, $S_L$, based on the relationships, is high quality.
Because clustering is unsupervised learning, we do not know which similarity matrix is high quality before the refinement process. Therefore, there is no procedure better than computing the weighted sum of $S_A$ and $S_L$ with the weight vector $\bm{\lambda}$ initialized to $(0.5,0.5)$, which results in a low-quality $S$.
However, if $\bm{\lambda}$ is adjusted to assign large weights to high-quality similarity matrices, we can obtain a high-quality similarity matrix by computing its weighted sum, as shown in the lower right of Fig.~\ref{lambda}.

The quality of the similarity matrices is increased by adjusting $\bm{\lambda}$ as mentioned above.
As mentioned in Section~\ref{3.3}, the similarity matrix $S$ is high quality when the graph corresponding to $S$ consists of $c$ connected components.
Therefore, SClump aims to refine $S$ to $S^c$, whose corresponding graph consists of $c$
connected components, by adjusting $\bm{\lambda}$.
To achieve this, SClump solves the following optimization problem, with variables $S^*$, $\bm{\lambda}$, and $X$:

\begin{align}
\label{eq_optS}
&(S^c, \bm{\lambda}) = \argmin_{S^*, \bm{\lambda}} \ \ ||S^*-S||_F^2 + \alpha||S^*||_F^2 + \beta||\bm{\lambda}||_2^2 ~ \notag \\ 
	&~~~~~~~~~~~~~~~~~~~~~~~~ +2\gamma ~tr(X^TL_sX), \\ 
&s.t. \ \ \Sigma_{j=1}^{n}S^*(i, j) =1, \  ||\bm{\lambda}||_1 = 1, \  S^*(i, j)\ge0,\ \lambda_{k}\ge0, \notag
\end{align}

\vspace{+1mm}
\noindent
where $X=(\bm{x}_1, \ldots, \bm{x}_c) \in \mathbb{R}^{n \times c}$ and each $\bm{x}_i$ is the eigenvector corresponding to the $i$-th smallest eigenvalue of the Laplacian matrices $L_s$.
Additionally, $\alpha$, $\beta$, and $\gamma$ are hyper parameters.
The objective function in Eq.~(\ref{eq_optS}) consists of four terms.
Its first term learns $S^*$, which approximates $S$ derived in Eq.~(\ref{eq_ini_HAN}).
The second and third terms are regularization terms, which prevent $S^*$ and $\bm{\lambda}$ from overfitting input data.
The fourth term optimizes $S$ to $S^*$ so that its corresponding graph consists of $c$ connected components.
This term achieve the refinement process as described in Section~\ref{3.3}.
Finally, the four constraints normalize $S^*$ and $\bm{\lambda}$, and make all their elements non-negative.

We solve Eq.~(\ref{eq_optS}) using an iterative update approach, called the coordinate descent approach~\cite{Li}.
In each iteration, two of the three variables are fixed, whereas the remaining variable is updated.
When the value of the objective function in Eq.~(\ref{eq_optS}) converges, we end the iterative update approach.

\vspace{+3mm}
\subsubsection*{Clusters with a High Mean Silhouette Score}
\label{4.3}

Additionally, we also aim to obtain clusters that have a high mean silhouette score, as mentioned in Section~\ref{3.3}.
We add a stopping criteria in the iterative update approach using the mean silhouette score \cite{FWang}.
This measure provides a silhouette score $sil(o_i)$ for each object $o_i$ in a cluster $C_h$, taking into account both condensability $a(o_i)$ and separability $b(o_i)$, as follows:

\begin{align}
\label{silhouette}
sil(o_i) &= \frac{b(o_i)-a(o_i)}{max \{a(o_i), b(o_i)\}}, \\ \notag
a(o_i) &= \frac{\sum_{j=1,j\ne i}^{n} w_{j,h} ~d(\bm{a}_i, \bm{a}_j)}{|C_h|-1}, \\ \notag
b(o_i) &= \min_{l \ne h} \frac{w_{j, l} ~d(\bm{a}_i, \bm{a}_j)} {|C_l|}, \notag
\end{align}
\noindent
where $w_{j,h}$ is the indicator, which equals 1 when $o_j$ is in the cluster $C_h$, and 0 otherwise.
Additionally, $d(\bm{a}_i, \bm{a}_j)$ is the squared Euclidean distance between $\bm{a}_i$ and $\bm{a}_j$.

The calculation of the silhouette score requires a set of clusters.
Therefore, we first obtain a set of clusters $\mathcal{C}_{iter}$ by inputting $S^*_{iter}$ into spectral clustering, where $S^*_{iter}$ is $S$ updated $iter$ times using the iterative update approach. 
Next, we compute the mean silhouette score $m_{sil}(iter)= \sum_{i=1}^{n}sil(o_i)/n$ based on the set of clusters $\mathcal{C}_{iter}$.
We terminate the iterative update when the $m_{sil}(iter-1)$ reaches the local maximum that satisfies $m_{sil}(iter-2) < m_{sil}(iter-1) > m_{sil}(iter)$. 
We output the set of clusters $\mathcal{C}_{iter-1}$ and achieve the clustering of the AN.
The pseudo code of our proposed method is summarized in Algorithm~1.

\begin{algorithm}
\caption{Clustering an AN using the refinement process}
\begin{algorithmic}
\STATE 	Input: An AN with $A$ and $L$, and a number of clusters $c$
\STATE 	Output: A set of clusters $\mathcal{C}$
\STATE ~1: Input all $\bm{a}_i$, which are updated using PCA, into Eq.~(\ref{s1}) \\~~~~to derive $S_A$
\STATE ~2: From the relationships in $L$, derive $S_{La}$ using Eq.~(\ref{s2}) \\~~~ and $S_{Lb}$ using Eq.~(\ref{s2_})
\STATE ~3: Normalize each matrix $S_k$, derived on Lines 1 and 2, \\~~~ using $\sum_{j=1}^{n}S_k(i,j)=1$
\STATE ~4: Derive a similarity matrix $S$ using Eq.~(\ref{eq_ini_HAN})
\STATE ~5: $S^*_0 \leftarrow S$
\STATE ~6: {\bf for} $iter \in \{0,1,\ldots,\infty \}$ {\bf do}
\STATE ~7: ~~~Obtain a set of clusters $\mathcal{C}_{iter}$ by inputting $S^*_{iter}$ \\~~~~~~ into spectral clustering
\STATE ~8: ~~~{\bf if} $m_{sil}(iter-2) < m_{sil}(iter-1) > m_{sil}(iter)$ {\bf do}
\STATE ~9: ~~~~~~$stop \leftarrow iter-1$
\STATE 10: ~~~~~~Output $\mathcal{C}_{stop}$
\STATE 11: ~~~{\bf if} Eq.~(\ref{eq_optS}) converges {\bf do}
\STATE 12: ~~~~~~$conv \leftarrow iter-1$
\STATE 13: ~~~~~~Output $\mathcal{C}_{conv}$
\STATE 14: ~~~Update $S^*_{iter}$ to $S^*_{iter+1}$ using the iterative update
\end{algorithmic}
\end{algorithm}


\vspace{+4mm}
\section{Evaluation Experiments}
\label{5}
We implemented Algorithm~1 in Python\footnote[5]{https://github.com/yururyuru00/SpectralClusteringwithRefine}.
Table~\ref{parameters} shows the datasets containing information on social networks\footnotemark[1] and citation networks\footnotemark[2] that were used for our experiments.

\begin{table}[b]
	\caption{Summary of the benchmark datasets}			
	\begin{center}
		\begin{tabular}{|c||r|r|r|} \hline
			Dataset & \#Objects & \#Dimensions of $\bm{a}_i$ & \#Clusters \\ \hline \hline
			football\footnotemark[1] & 248 & 11,806 &20\\ \hline
			politics-uk\footnotemark[1] & 419 & 19,868 & 5\\ \hline
			olympics\footnotemark[1] & 464 & 18,455 & 28\\ \hline
			cora\footnotemark[2] & 2,708 & 1,433 & 7\\ \hline
			citeseer\footnotemark[2] & 3,312 & 3,703 & 6\\
\hline
 		\end{tabular}
	\end{center}
	\label{parameters}
\end{table}

\subsubsection*{Experimental Settings}
On Line~1 of Algorithm~1, we derived a $S_A$ from attribute vectors $\bm{a}_i$.
In the case of citation networks, we derived $S_A$ using cosine similarity.

When solving Eq.~(\ref{eq_optS}) and using spectral clustering, it is necessary to solve the eigenvalue problems of Laplacian matrices derived from the similarity matrices.
The problems are relaxed normalized cut~(NCut) and ratio cut~(RCut) problems when the Laplacian matrices are normalized and unnormalized, respectively \cite{Luxburg}.
Therefore, we clustered the datasets using both normalized and unnormalized matrices.

As mentioned in Section~\ref{4.3}, we solve Eq.~(\ref{eq_optS}) using the iterative update approach to achieve the refinement process.
A major computation performed by the iterative update approach is the updating of $S^*=(\bm{s}_1, \dots, \bm{s}_n)$, which requires the solution of a quadratic programming problem for each $\bm{s}_i$.
Although the interior point method is the common algorithm used to solve the problem,
the time complexity for solving the problem for each $s_i$ is $\mathcal{O}(n^3)$. However,
because a fast algorithm, called simplex sparse representation, with complexity $\mathcal{O}(n^2)$ was proposed in \cite{Huang}, we used this algorithm.
Additionally, we updated the top
$m$ elements with a large similarity among $\bm{s}_i=(s_{i1}, \dots , s_{in})$,
and set the others to 0, as suggested by  \cite{Nie}. The time
complexity was thereby reduced significantly to make the
algorithm easy to apply to large-scale datasets.

\subsubsection*{Parameters Settings}
We set $\sigma$ in Eq.~(\ref{s1}) to~3 for football,~0.2 for politics-uk, and~0.1 for olympics.
We set the parameters in Eq.~(\ref{eq_optS}) based on the original papers \cite{Li}\cite{Nie}, which proposed Eq.~(\ref{eq_optS}).

\subsubsection*{Metrics for Evaluating Clustering Performance}
To evaluate the clustering accuracy, we used two popular metrics: normalized mutual information~(NMI) and purity \cite{Rezaei}.
The values of the two metrics range from 0 to 1, and the larger the value, the higher the clustering accuracy.

\subsection{Effect of Evaluating Transitive Relationships}
\label{5.3}

We proposed a process that evaluates transitive relationships, in addition to direct relationships.
To verify the effect of the process, we changed $\theta$ and fixed $\delta$.
When $\theta=1$, only direct relationships were evaluated; and
when $\theta>1$, transitive relationships that corresponded to the SPs with a length up to $\theta$ ware also evaluated.

Table~\ref{s2_ari} shows the NMI when we executed Algorithm~1 on the datasets while varying $\theta$.
The NMI was the best for all the datasets when we set $\theta$ to~3.
Therefore, these results demonstrate that evaluating both direct and transitive relationships was effective for clustering ANs.

\begin{table}[t]
	\caption{Clustering accuracy~(NMI) for all the datasets}
	\label{s2_ari}
	\begin{center}
		\begin{tabular}{|c||c||cccc|} \hline 
			& \multicolumn{1}{|c||}{Direct} & \multicolumn{4}{|c|}{Direct + transitive}\\
			Dataset & $\theta=1$ & $\theta=2$ & $\theta=3$ & $\theta=4$ & $\theta=5$\\ \hline \hline
   		  	football & ~~~0.87~~~ & ~~\bf{0.92}~~ & ~~\bf{0.92}~~ & ~~\bf{0.92}~~ & ~~\bf{0.92}~~ \\ \hline
			cora & 0.39 & 0.41 & \bf{0.44} & 0.43 & \bf{0.44} \\ \hline
     			politics-uk & 0.89 & 0.91 & \bf{0.93} & \bf{0.93} & \bf{0.93} \\ \hline
     			olympics & 0.92 & 0.92 & \bf{0.93} & 0.92 & 0.92 \\ \hline
			citeseer & 0.38 & \bf{0.39} & \bf{0.39} & \bf{0.39} & \bf{0.39} \\ \hline
          \end{tabular}
	\end{center}
\end{table}

\subsection{Effect of the Refinement of the Similarity Matrices in ANs}
\label{5.4}

We proposed a process for clustering ANs using the refinement, to increase the clustering accuracy.
To verify whether the refinement process increases the clustering accuracy, we executed Algorithm~1 on the experimental datasets.

\begin{figure*}[b]
\begin{center}
\includegraphics[width=17cm, height=9cm]{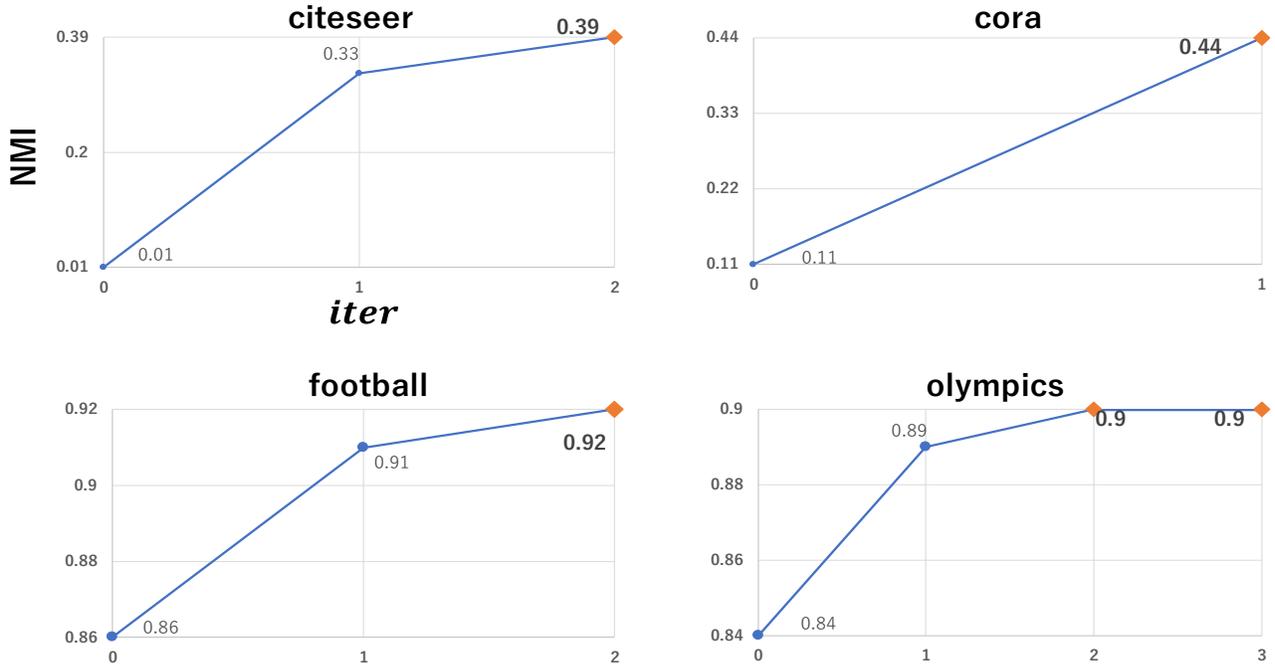}
\vspace{-3mm}
\end{center}
\caption{Effect of the refinement process at each iteration of Algorithm~1}
\label{experiment}
\end{figure*}

Figure~\ref{experiment} shows the NMI of clustering the datasets using the refinement process.
The horizontal axes show $iter$, which is the number of iterative updates of Eq.~(\ref{eq_optS}) required to achieve the refinement process. The vertical axes show the NMI when inputting $S^*_{iter}$ updated $iter$ times into spectral clustering, where $0\le iter \le stop$ and $stop$ refers to $stop$ in the pseudo code of Algorithm~1.
Here $S^*_0$ is the unrefined $S$, and each $S^*_{iter}~(0<iter\le stop)$ is a partially refined similarity matrix\footnote[6]{We use a term ``partially'' to indicate that that matrix is not minimized until the convergence of Eq.~(\ref{eq_optS}) and its minimization is terminated by the stopping criteria.}.

For all the datasets, Fig.~\ref{experiment} shows that the NMI of the partially refined $S^*_{stop}$ was greater than the NMI of the unrefined $S^*_0$.
In particular, for the citeseer and cora datasets, we confirmed that the partial refinement increased the NMI.
Therefore, these results demonstrate that the partial refinement, which updates $S$ until the stopping criteria was satisfied, increased the clustering accuracy for all the datasets.

\begin{figure}[!t]
\centering
\includegraphics[height=5cm, width=8.4cm]{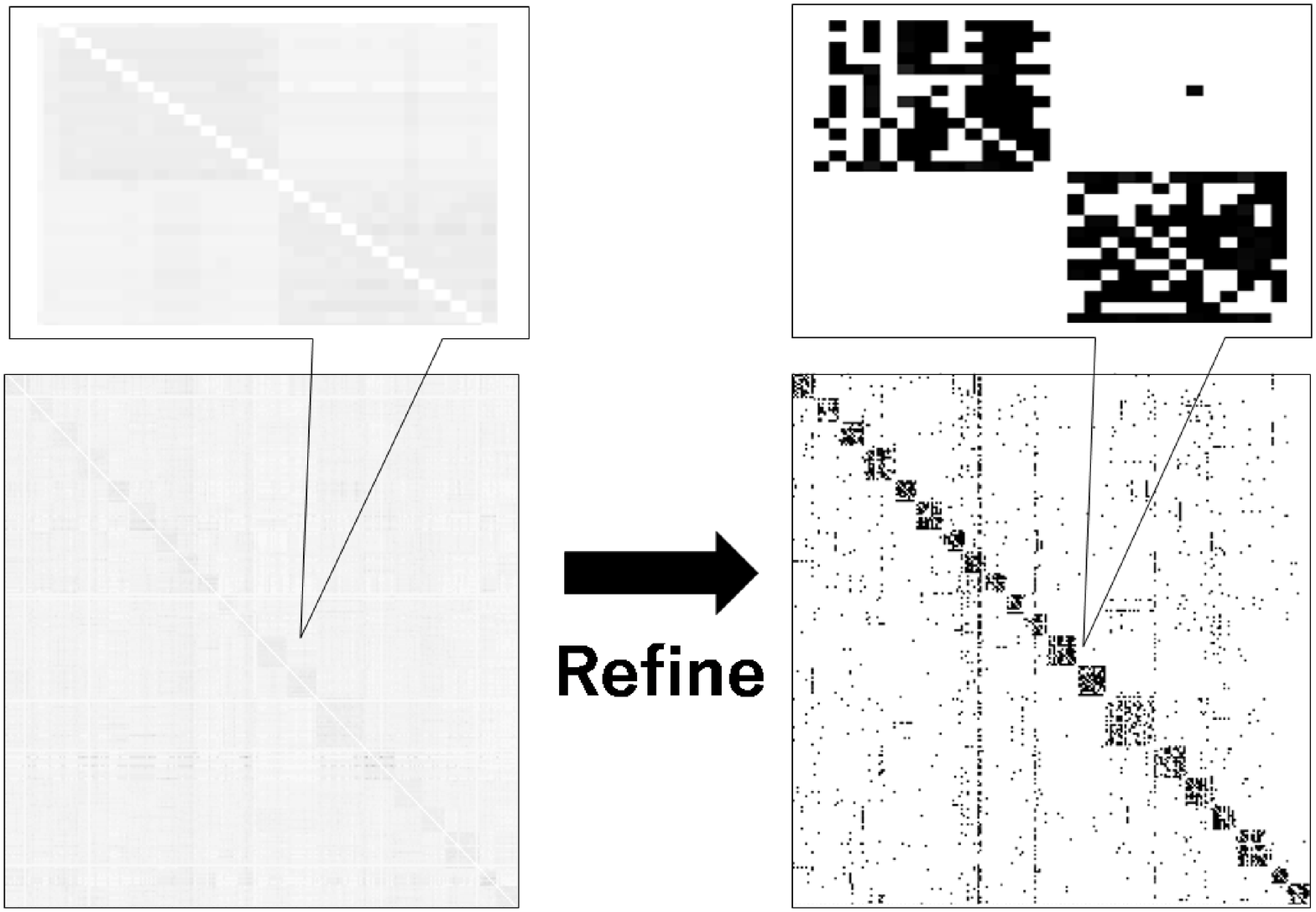}
\hspace{+10mm} football
\\
\includegraphics[height=5cm, width=8.4cm]{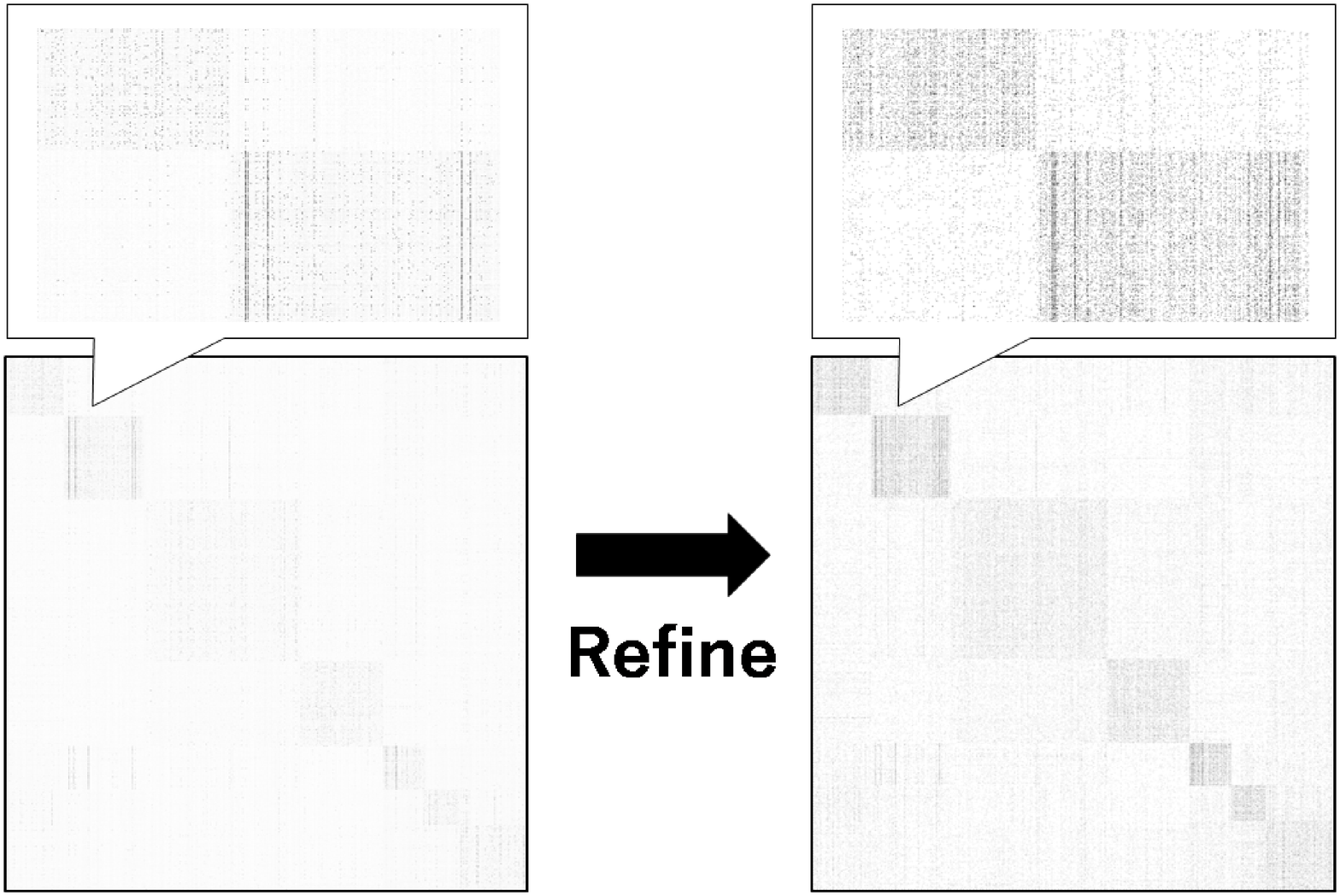}
\hspace{+20mm} cora
\caption{Visualization of similarity matrices before and after the partial refinement}
\label{refining_S}
\end{figure}

Additionally, to verify whether the partial refinement actually increases the quality of the similarity matrix, Fig.~\ref{refining_S} visualizes the unrefined $S$~(left-hand side) and the partially refined $S^*_{stop}$~(right-hand side) for the football and cora datasets.
Comparing the enlarged $S$ with the enlarged $S^*_{stop}$, it is clear that $S^*_{stop}$ is almost the same as the high-quality similarity matrix shown in the lower right of Fig.~\ref{highqualityS}. 
Each diagonal block in $S^*_{stop}$ indicates similarities in a connected component in $G(S^*_{stop})$ similar to one of the supervised clusters.
We confirmed this finding in all the experimental datasets.
Therefore, these results demonstrate that refinement increased the quality of the similarity matrix for all the datasets.

We also compared our clustering method with three state-of-the-art clustering methods without using the refinement process: ~MVAP~\cite{Cwang}, CESNA \cite{JYang}, and SSB\cite{Chen}.
Regarding the clustering accuracy of the comparative studies, we referred to \cite{Cwang}\cite{Chen}.
Table~\ref{compare} shows the NMI and purity of the four clustering methods: MVAP, CESNA, SSB, and our method~(Algorithm~1) with the refinement process. For this experiment, we executed Algorithm~1 for both normalized and unnormalized Laplacian matrices.
Table~\ref{compare} shows that our method outperformed the state-of-the-art clustering methods.
The results confirm that our method using the partial refinement of the similarity matrix was very effective for high-accuracy clustering ANs for the olympics, cora, and citeseer datasets.

\begin{table*}[b]
	\centering
	\caption{Results of the experiments compared with existing methods}
	\begin{center}
		\begin{tabular}{|cc||ccc||cc|}
		\hline
		& & \multicolumn{3}{|c||}{No refinement} & \multicolumn{2}{|c|}{Refinement~(Algorithm~1)}\\
		Dataset~&~Metric~&~MVAP~&~CESNA~&~SSB~&~Normalized~&~Unnormalized~ \\ \hline
		\multirow{2}{*}{olympics} & NMI & Not listed & 0.48 & 0.50 & 0.90 & \bf{0.91}\\
		                                & Purity & Not listed & 0.36 & 0.36 & 0.93 & \bf{0.94}\\
		\hline
		\multirow{2}{*}{cora} & NMI & 0.26 & 0.12 & 0.20 & \bf{0.44} & 0.38\\
		                            & Purity & 0.27 & 0.30 & 0.38 & \bf{0.60} & 0.54\\
 		\hline
		\multirow{2}{*}{citeseer} & NMI & 0.21 & 0.18 & 0.23 & 0.31 & \bf{0.39}\\
		                                & Purity & 0.26 & 0.19 & 0.40 & 0.51 & \bf{0.61}\\
		\hline
		\end{tabular}
	\end{center}
	\label{compare}
\end{table*}

\subsection{Refinement Details}
\label{5.5}

\begin{figure*}[t]
\begin{center}
\includegraphics[width=16cm, height=9.6cm]{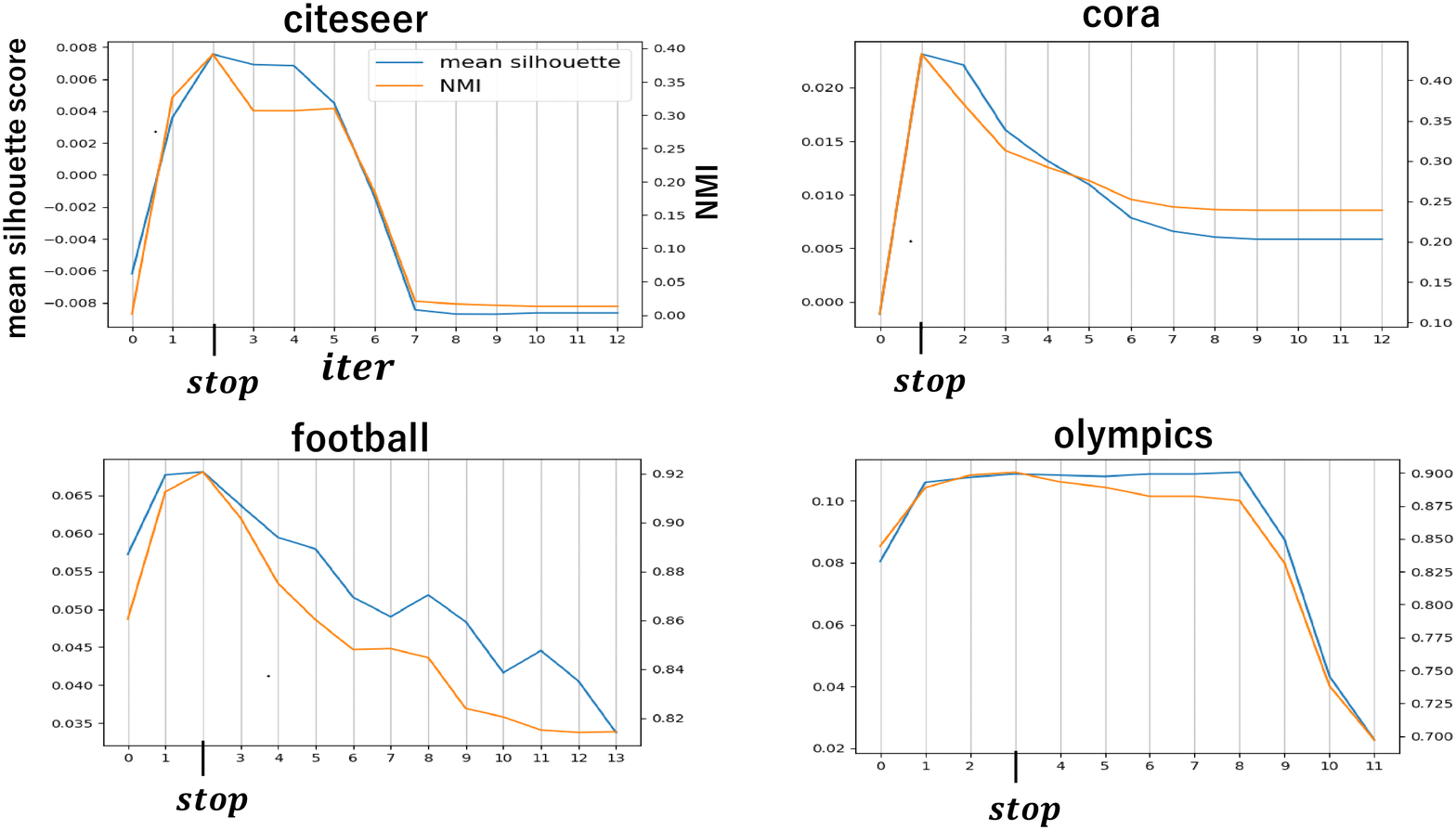}
\end{center}
\vspace{-3mm}
\caption{Silhouette scores and NMIs at each iteration of Algorithm~1 for the datasets}
\label{silhouette_score}
\end{figure*}

As shown in Fig.~\ref{experiment}, we confirm that the partial refinement, which updated $S$ until satisfying the stopping criteria, increased the NMI.
By contrast, we found that the complete refinement of the similarity matrix, which means minimizing Eq.~(\ref{eq_optS}) without the stopping criteria of the mean silhouette score, reduced the NMI.
To verify this finding, the blue lines in Fig.~\ref{silhouette_score} show the mean silhouette scores.
The horizontal axes show $iter$, which is the number of iterative updates of the refinement process, where $0\le iter \le conv$, and $conv$ refers to $conv$ in the pseudo code of Algorithm~1.
The blue lines show the mean silhouette scores based on the predictive clusters $\mathcal{C}_{iter}$, which were obtained at each iteration.

For all the datasets, Fig.~\ref{silhouette_score} shows that the mean silhouette scores reached the maximum at $stop$, and thereafter reduced when $iter$ became larger than $stop$.
The effect of the refinement was to optimize the similarity matrix $S$ so that $G(S)$ consisted of $c$ connected components.
Additionally, the mean silhouette score was high when the sizes of connected components obtained as clusters were almost the same and separate from each other.
Therefore, the high silhouette score at $stop$ means that the partial refinement separated connected components whose cluster sizes were almost the same.
Moreover, for all the datasets in Fig.~\ref{silhouette_score}, there were strong correlations between the mean silhouette scores (blue lines) and NMIs (orange lines).
Therefore, we confirmed that the unevenly sized and non-separated connected components generated by the complete refinement beyond the stopping criteria led to a reduction of the clustering accuracy.

\section{Discussion of Related Work}
\label{6}
\vspace{+1mm}

In this section, we discuss the novelty of our study when compared with SClump \cite{Li}, which is the basis of our study.
As mentioned in Section~\ref{3.4}, SClump is applicable to heterogeneous information networks~(HINs) in which ``each object has only a single attribute of a discrete value,'' which is not true of our ANs in which ``each object has multiple attributes of continuous values."
We explain why SClump cannot be applied to ANs.

As mentioned in Section \ref{4.3}, the refinement of the similarity matrix $S$ requires multiple similarity matrices summarized in $S$.
To compute each similarity of the matrices, SClump uses a meta-path, which is a measure for calculating similarity based on a single attribute that each object has and relationships between objects in HINs.
Therefore, for the clustering of ANs with the refinement process, it is necessary to propose a new measure based on multiple attributes of each object and relationships between objects in ANs.
We proposed new measures in Section~\ref{4.2}, and these measures enabled the refinement to be applicable to not only HINs but also ANs.

Additionally, regarding the refinement process, the authors of SClump proposed the complete refinement which updates $S$ until Eq.~(\ref{eq_optS}) converges.
By contrast, we proposed the partial refinement, in addition to the traditional complete refinement, by adding a stopping criteria in Eq.~(\ref{eq_optS}).
The evaluation experiments in Section~\ref{5.4} show that the partial refinement is effective in the clustering of ANs.
Furthermore, in Section~\ref{5.5}, we verified that the complete refinement reduced the clustering accuracy of ANs.


\section{Conclusion}
\label{7}

We proposed a method for high-accuracy clustering of ANs.
The method contains the process to evaluate both direct and transitive relationships between objects to derive a high-quality similarity matrix.
In addition, it contains another process to apply refinement to ANs to increases the clustering accuracy.

Evaluation experiments showed that the first process is effective for clustering ANs.
We also confirmed that partial refinement, rather than complete refinement proposed in \cite{Li}, increases the clustering accuracy of ANs.
We showed that our proposed method outperforms state-of-the-art clustering methods without the refinement.
From the above, we confirmed that our proposed method is effective for the high-accuracy clustering of ANs.




\begin{thebibliography}{00}

\bibitem{Khan}
Khan N, Yaqoob I, Hashem IA, Inayat Z, Ali M, Kamaleldin W, Alam M, Shiraz M, Gani A. Big data: survey, technologies, opportunities, and challenges. The scientific world journal. 2014;2014.

\bibitem{Bothorel}
Bothorel C, Cruz JD, Magnani M, Micenkova B. Clustering attributed graphs: models, measures and methods. Network Science. 2015 Sep;3(3):408-44.

\bibitem{Zhiqiang}
Xu Z, Ke Y, Wang Y, Cheng H, Cheng J. A model-based approach to attributed graph clustering. InProceedings of the 2012 ACM SIGMOD international conference on management of data 2012 May 20 (pp. 505-516).

\bibitem{Zhiqiang2}
Xu Z, Ke Y, Wang Y, Cheng H, Cheng J. GBAGC: A general bayesian framework for attributed graph clustering. ACM Transactions on Knowledge Discovery from Data (TKDD). 2014 Aug 25;9(1):1-43.

\bibitem{Luxburg}
Von Luxburg U. A tutorial on spectral clustering. Statistics and computing. 2007 Dec 1;17(4):395-416.

\bibitem{FWang}
Wang F, Franco-Penya HH, Kelleher JD, Pugh J, Ross R. An analysis of the application of simplified silhouette to the evaluation of k-means clustering validity. InInternational Conference on Machine Learning and Data Mining in Pattern Recognition 2017 Jul 15 (pp. 291-305). Springer, Cham.

\bibitem{Li}
Li X, Kao B, Ren Z, Yin D. Spectral clustering in heterogeneous information networks. InProceedings of the AAAI Conference on Artificial Intelligence 2019 Jul 17 (Vol. 33, pp. 4221-4228).

\bibitem{Huang}
Huang J, Nie F, Huang H. A new simplex sparse learning model to measure data similarity for clustering. InTwenty-Fourth International Joint Conference on Artificial Intelligence 2015 Jun 27.

\bibitem{Nie}
Nie F, Wang X, Jordan MI, Huang H. The constrained laplacian rank algorithm for graph-based clustering. InThirtieth AAAI Conference on Artificial Intelligence 2016 Mar 2.

\bibitem{Rezaei}
Rezaei M, Fr\"{a}nti P. Set matching measures for external cluster validity. IEEE Transactions on Knowledge and Data Engineering. 2016 Apr 7;28(8):2173-86.

\bibitem{Cwang}
Wang CD, Lai JH, Philip SY. Multi-view clustering based on belief propagation. IEEE Transactions on Knowledge and Data Engineering. 2015 Nov 25;28(4):1007-21.

\bibitem{JYang}
Yang J, McAuley J, Leskovec J. Community detection in networks with node attributes. In2013 IEEE 13th International Conference on Data Mining 2013 Dec 7 (pp. 1151-1156). IEEE.

\bibitem{Chen}
Chen H, Yu Z, Yang Q, Shao J. Attributed Graph Clustering with Subspace Stochastic Block Model. Information Sciences. 2020 May 21.

\end{thebibliography}
\end{document}